\documentclass[onecolumn,amsmath,amssymb,nofootinbib]{qm2008_abs}
\usepackage{graphicx}
\usepackage{dcolumn}
\usepackage{bm}
\begin{document}

\title{{\Large Vector-axialvector mixing in hot matter
\\
and its hadronic effective field theory description}}

\bigskip
\bigskip
\author{\large C.~Sasaki$^1$}
\email{csasaki@ph.tum.de}
\author{\large M.~Harada$^2$}
\author{\large W.~Weise$^1$}

\affiliation{$^1$ Physik-Department,
Technische Universit\"{a}t M\"{u}nchen,
D-85747 Garching, Germany}
\affiliation{$^2$ Department of Physics, 
Nagoya University,
Nagoya, 464-8602, Japan}
\bigskip
\bigskip

\begin{abstract}
\leftskip1.0cm
\rightskip1.0cm
We discuss the importance of the axialvector meson in chiral
phase transition at finite temperature.
We show that there exists a significant contribution in the
vector spectral function from the axialvector meson
through the vector-axialvector mixing in hot matter.
\end{abstract}

\maketitle

\section{Introduction}

Changes of hadron properties are expected to be indications 
of the tendency towards chiral symmetry restoration in hot 
and/or dense QCD. 
In particular, the short-lived vector mesons like the $\rho$ 
mesons are expected to carry information on the modifications 
of hadrons in matter~\cite{review}. In the presence of hot matter 
the vector and axialvector current correlators are mixed due 
to the pion in the heat bath, and a process is described in a 
model-independent way at low temperatures as a low-energy 
theorem of chiral symmetry~\cite{theorem}. The vector spectral
function is then modified by axialvector mesons through the 
mixing theorem~\cite{vamix,UBW}.

The validity of the theorem is, however, limited to temperatures 
$T < 2f_\pi$, where $f_\pi$ is the pion decay constant in vacuum.
At higher temperatures hadrons other than pions are thermally 
activated. Thus one needs in-medium correlators 
systematically involving those excitations. In this contribution 
we show the effects of the mixing (hereafter V-A mixing) and how the 
axialvector mesons affect the spectral function near the chiral 
phase transition within an effective field theory.

\section{Role of axialvector mesons}

Several models exist which explicitly include the axialvector 
meson in addition to the pion and vector meson consistently with 
the chiral symmetry of QCD, such as the Massive Yang-Mills 
theory ~\cite{MassiveYM}, the anti-symmetric tensor field 
method~\cite{anti:tensor} and the model based on the generalized 
hidden local symmetry (GHLS)~\cite{ghls:pr,Kaiser:1990yf}.
These models are equivalent~\cite{Kaiser:1990yf,equi} for 
tree-level amplitudes in low-energy limit.
Recently a systematic perturbation scheme based on the GHLS has 
been constructed~\cite{ghls,kek}. In this approach the Weinberg
sum rules~\cite{Weinberg} are stable against the renormalization 
group evolution at one-loop~\cite{ghls}~\footnote{
 The GHLS Lagrangian does not include $\bar{q}q$ scalar modes
 which are assumed to be heavier than other mesons incorporated.
 This may not be true near the critical point within the 
 Ginzburg-Landau picture of the phase transition. The scalar
 mesons thus modify the renormalization group structure.
}.

The critical temperature is defined as the temperature at which
the vector and axialvector current correlators coincide.
When these correlators are saturated by the lowest lying meson,
we have~\cite{ghls}
\begin{equation}
G_A - G_V \propto M_\rho^2 (M_{a_1}^2 - M_\rho^2) 
= M_\rho^2 \delta M^2\,.
\end{equation}
Then the chiral symmetry restoration implies that
either $\delta M = 0$ or $M_\rho = 0$ (or both):
Either the $\rho$-$a_1$ mass difference $\delta M$ or
the $\rho$ meson mass is identified
with a measure
of the spontaneous chiral symmetry breaking and acts as an order
parameter of the chiral phase transition. We consider $\delta M$
changing with temperature intrinsically such that one has
$G_A - G_V = 0$ at the chiral transition.

Thus, the {\it bare }axialvector meson mass 
depends on temperature as
\begin{equation}
M_{a_1}^2 = M_\rho^2 + \delta M^2(T)\,.
\end{equation}
Note that the dropping masses following the Brown-Rho 
scaling~\cite{br} are achieved if the bare $\rho$ mass goes to
zero at the critical point, which is indeed the case in the
vector manifestation (VM) scenario~\cite{vm}.

\begin{figure}
\begin{center}
\includegraphics[width=10cm]{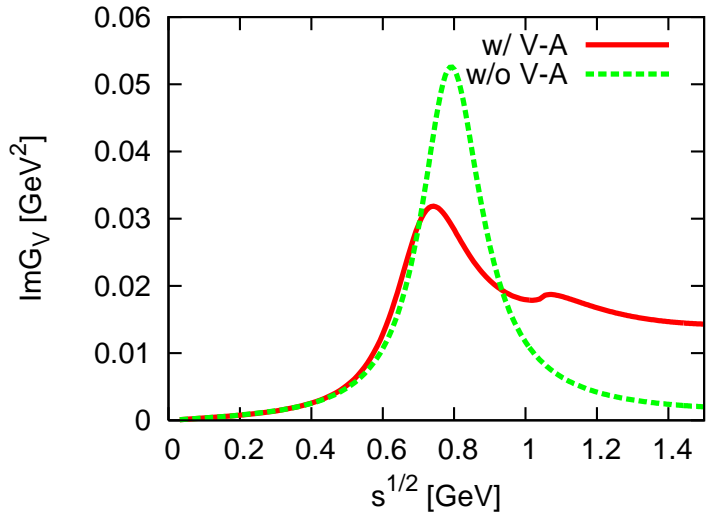}
\caption{
The vector spectral function at temperature $T/T_c = 0.85$
with the critical temperature $T_c = 200$ MeV, calculated
in the $\rho$-meson rest frame. 
The solid curve is obtained in the full calculation. The dashed 
line is calculated eliminating the axialvector meson from the theory.
}
\label{vamix}
\end{center}
\end{figure}

\begin{figure}
\begin{center}
\includegraphics[width=10cm]{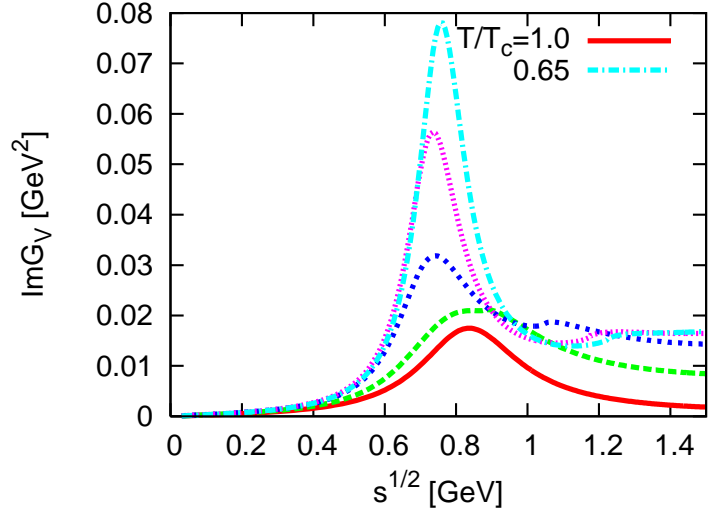}
\caption{
The vector spectral function at several temperatures 
$T/T_c = 0.65$-$1.0$ in steps of $0.1$
(from top to bottom).
}
\label{tdep}
\end{center}
\end{figure}

We show the spectral function of the vector current correlator
calculated in the GHLS theory in Fig.~\ref{vamix}.
Two cases are compared; one includes the V-A mixing and the other
does not. The spectral function has a peak at $M_\rho$ and a 
bump around $M_{a_1}$ due to the mixing. The height of the
spectrum at $M_\rho$
is enhanced and a contribution above $\sim 1$ GeV is gone when
one omits the $a_1$ in the calculation.
In Fig.~\ref{tdep} we show the temperature dependence of the
vector spectral function.
One observes a systematic downward shift of the enhancement
around the $a_1$ mass with temperature,
and $M_{a_1}$ comes together with $M_\rho$ at $T_c$. The V-A
mixing eventually vanishes there which is a direct consequence
of vanishing coupling of $a_1$ to $\rho$-$\pi$. This feature is 
unchanged even if an explicit scalar field is present~\cite{hsw}.

\begin{figure}
\begin{center}
\includegraphics[width=10cm]{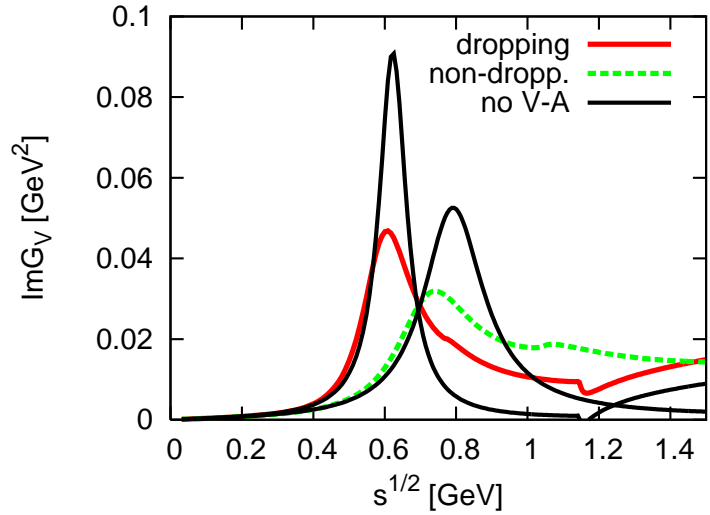}
\caption{ 
 The vector spectral function at temperature $T/T_c = 0.85$
 with dropping (solid) and non-dropping (dashed) $\rho$ bare 
 mass. The black solid lines indicate the results eliminating
 the $a_1$ mesons.
}
\label{br}
\end{center}
\end{figure}

In case of dropping $\rho$ and $a_1$ masses,
the spectral function is enhanced compared to that without mass 
dropping since the $\rho$ decay width is narrower~\cite{hls:dl}
as shown in Fig.~\ref{br}.
The feature that the $a_1$ meson suppresses the spectral function
through the V-A mixing is unchanged.
It should be noted that the vector meson becomes the chiral partner
of the pion and vector meson dominance is strongly violated
when the chiral symmetry is restored in the VM~\cite{vm}. 
This induces a significant reduction of the vector spectral 
function~\cite{hls:dl,bhhrs}. On the other hand, the pion form
factor is still vector-meson dominated at $T_c$ if the dropping 
$\rho$ and $a_1$ join in the same chiral multiplet~\cite{ghls}.

\section{Conclusions}

We have performed a systematic study of the V-A mixing in the current
correlation functions and its evolution with temperature.
The axialvector meson contributes significantly to 
the vector spectral function; the presence of the $a_1$ 
reduces the vector spectrum around $M_\rho$
and enhances it around $M_{a_1}$.

One interesting application of this thermal spectral function
is to study dilepton production in relativistic heavy-ion collisions.
The change of the $a_1$-meson properties and V-A
mixing near the critical point has not been properly treated so far
in dilepton processes in the context of the chiral phase transition.
In order to deal with the dileptons one needs to account for
other collective excitations and many-body interactions 
as well as the time evolution of a created fireball~\cite{dileptons}. 
Such effects can screen signals of chiral restoration~\cite{bhhrs} 
and make an interpretation of broad in-medium
spectral functions in terms of a changing chiral order parameter
quite difficult~\cite{qcdsr}.
The situation at RHIC and/or LHC might be very different from
SPS. At SPS energies many-body effects come from
the presence of baryons. These effects are expected to be
much reduced in very
hot matter with relatively low baryon density. 
The present study may then be of some relevance for the high
temperature, low baryon density scenarios encountered
at RHIC and LHC.

\subsection*{Acknowledgments}

The work supported in part by BMBF and by the DFG cluster 
of excellence ``Origin and Structure of the Universe''.

\vspace*{0.3cm}

\end{document}